\documentclass[a4paper,12pt]{article}
\usepackage{amssymb}
\begin{document}
\title{Surprises in Bose-Einstein correlations}
\author{Kacper Zalewski\thanks{Supported in part by the KBN grant
2P03B 093 22}\\
M. Smoluchowski Institute of Physics, Jagellonian University\\ and\\ Institute of
Nuclear Physics, Krak\'ow, Poland}
\maketitle
\begin{abstract}
Four experimental results, which seem to contradict the established ideas about the
Bose-Einstein correlations in multiple particle production precesses, are briefly
presented and discussed.
\end{abstract}

\section{INTRODUCTION}

The study of Bose-Einstein correlations (further quoted BEC ) in multiple particle
production processes is an important source of information  about the interaction
region, i.e. about the region, where the hadrons are produced. One would like to know
the size and shape of this region as well as its orientation with respect to the
momentum of the incident particle and to the impact parameter vector. For the matter
inside one would like to know: its flows, its equation of state and its phase
transitions. Much information has already been obtained (cf. the reviews \cite{WIH},
\cite{WEI}, \cite{HEJ}, \cite{ZAL1} and references quoted there), its reliability,
however, depends on the correctness of our interpretation of the observed BEC.

Four surprising experimental results concerning BEC will be here briefly presented and
discussed. By surprising we do not just mean that they disagree with some specific
model, but that they seem to contradict the basic physical pictures used to build most
of the currently popular models. Perhaps in the future some trivial explanations will
be found, but if not, these observations may lead to a reinterpretation of the BEC
data and consequently to changes in our conclusions concerning the interaction region.
Thus the questions raised are interesting and potentially important.

\section{No inter-W BEC at LEP}
At LEP2 the reactions

\begin{eqnarray}\label{}
  e^+e^- \rightarrow  W^+W^- \rightarrow & q\overline{q}l\nu\\
  e^+e^- \rightarrow  W^+W^- \rightarrow & q\overline{q}q\overline{q}
\end{eqnarray}
have been observed. In the first, further denoted $W$, one of the $W$-bosons decays
into hadrons and the other one into leptons. In the second, further denoted $WW$, both
$W$-s decay into hadrons. Assuming that in reaction $WW$ there are no inter-$W$
correlations, it is a standard exercise in the calculus of probabilities to express
any two-particle distribution for reaction $WW$ in terms of the corresponding
single-particle and two-particle distributions for reaction $W$. One finds \cite{CDK}

\begin{equation}\label{}
\rho_2^{(WW)}(p_1,p_2) = 2\rho_2^{(W)}(p_1,p_2) + 2\rho_1^{(W)}(p_1)\rho_1^{(W)}(p_2).
\end{equation}
The surprise is that this formula works -- there is no evidence for inter-$W$
correlations \cite{ABR},\cite{ABB},\cite{BAR},\cite{ACC},\cite{ACH}.

There is a number of known factors, which reduce the inter-$W$ BEC (cf. e.g.
\cite{FIW}). The two $W$-s decay in different places and have non-zero relative
velocity. The partons from one $W$ can form hadrons with the partons from the other.
This effect -- known as color reconnection -- shortens the chain of interactions
leading to the final hadrons and consequently reduces the multiplicity of the final
particles. Since BEC increase the particle multiplicity, the two effects tend to
cancel. Nevertheless, the observed reduction of the inter-$W$ BEC seems to be much
stronger than what can be easily understood.

One can try various explanations of this fact. The belief that BEC for identical
bosons must lead to attraction in momentum space is based on a restricted class of
models, where the two-particle density matrix in the momentum representation is a
symmetrized product of single particle density matrices \cite{KAR},\cite{ZAL2}. This
does not have to be the case. For instance the amplitude

\begin{equation}\label{}
  \psi(p_1,p_2) = Sign\left[(p_1-p_2)(x_1-x_2)  \right]
  \times\left(e^{ip_1x_1+ip_2x_2} - e^{ip_2x_1+ip_1x_2}\right)
\end{equation}
has the correct symmetry, but corresponds to a different density matrix and leads to
repulsion in momentum space instead of the attraction. Perhaps one should look for
more general density matrices than those currently used. H\"{a}kkinen and Ringn\'er
\cite{HAR} suggested that the absence of the inter-$W$ correlations is natural in the
framework of string models and postulated it. This point of view was supported by Bo
Andersson (cf. e.g. \cite{AND}). Up to now, however, no convincing motivation for it
has been published.

\section{The Bertsch-Pratt radii}

The next three surprising results have come from the study of heavy ion collisions at
RHIC. In order to present them it is convenient to introduce the Bertsch-Pratt
coordinate frame \cite{GRA}, \cite{PRA}, \cite{BER}. This is a Cartesian frame with
its axis $L$ -- for longitudinal -- along the event axis. For heavy ion collisions the
event axis is along the direction of the momentum difference of the two nuclei in the
centre of mass system or equivalently in the laboratory system. The other two axes are
in the plane perpendicular to $L$ and are defined separately for each pair of
particles. Denoting the four-momenta of the two particles forming the pair by $p_1$
and $p_2$ let us define two more four-vectors

\begin{equation}\label{}
  K = \frac{1}{2}(p_1 + p_2); \qquad q = p_1 - p_2.
\end{equation}
The second axis of the Bertsch-Pratt frame -- denoted $out$ -- is parallel to the
transverse (with respect to $L$) component of $\textbf{K}$. The third axis -- denoted
$side$ -- is perpendicular to the $L$ and $out$ axes, or equivalently to the $L$-axis
and to the $\textbf{K}$ vector.

The correlation function measured in experiment, after a number of corrections which
will not be discussed here, is parameterized as

\begin{equation}\label{berpra}
C(K,q) = 1 + \lambda  e^{-R_L^2(K_T)q_L^2 - R_{o}^2(K_T)q_{o}^2 -
R_{s}^2(K_T)q_{s}^2}.
\end{equation}
The coefficient $\lambda$ is a constant and $R_i(K_T)$ are the three Bertsch-Pratt
radii. The following three assumptions had been accepted.
\begin{itemize}
  \item The momenta $\textbf{p}_i$ are measured in the local co-moving system (LCMS) obtained
  from the laboratory (or cms.)  system by a Lorentz transformation along the $L$
  direction such that in the LCMS: $K_L = 0$.
  \item The distribution of momenta $\textbf{p}_i$ is invariant with respect to boosts along
  the $L$ axis. This is known as boost invariance \cite{BJO}.
  \item The distribution of pairs of momenta $\textbf{p}_1,\textbf{p}_2$ does not change under rotations
  around the $L$ axis. This means either that the collisions are central, or that the
  data is averaged over the angle between $\textbf{K}_T$ and the impact parameter vector
  $\textbf{b}$.
\end{itemize}
Under these assumptions vector $\textbf{K}$ in the arguments of the $R_i(K)$ can be
replaced by the length of its transverse component $K_T = |\textbf{K}_T|$. The
dependence on the time component of $K$, which is allowed, is not written explicitly.
Let us consider now the three surprising results referred to collectively as the RHIC
(BEC) puzzle.

\section{Approximate equality $R_s(K_T) \approx R_o(K_T)$}

At $\textbf{K}_T = 0$ each direction perpendicular to $L$ can be considered parallel
as well as perpendicular to $\textbf{K}_T$ and consequently

\begin{equation}\label{}
  R_o(0) = R_s(0).
\end{equation}
The relation of these parameters to the space distribution of the sources of pions is

\begin{eqnarray}\label{}
  R_o^2(0) = \langle \tilde{x}_{out}^2 \rangle_0,\\
  R_s^2(0) = \langle \tilde{x}_{side}^2 \rangle_0,
\end{eqnarray}
where $\tilde{x} = x - \overline{x}$ is the deviation of the position four-vector of
the source from a fixed space time point $\overline{x}$ corresponding to the average
position of the source in space-time. The averaging is over all the pairs of
particles, which have $K_T = 0$ and are of the type considered, say $\pi^+$. We use
the notation $\tilde{x} = \{\tilde{x}_L, \tilde{x}_{out},\tilde{x}_{side}, \tilde{t}
)$. For reasons discussed below models suggest that the ratio $R_s(K_T)/R_o(K_T)$
should increase with increasing $K_T$ \cite{ADL}, \cite{ADC}. It came, therefore, as a
surprise that this ratio remains approximately equal one, when $K_T$ increases from
zero \cite{ADL}, \cite{ADC}, \cite{MAN}. What is more, subsequent data indicate that
the ratio $R_o(K_T)/R_s(K_T)$ decreases below one with increasing $K_T$ \cite{ADL},
\cite{ENO}, \cite{LOP} as well as with decreasing impact parameter of the collision
\cite{ENO}.

The transition from $K_T = 0$ to $K_T \neq 0$ corresponds to a Lorentz transformation
with velocity

\begin{equation}\label{}
  \beta_T = \frac{K_T}{K_0}
\end{equation}
along the direction of $\textbf{K}_T$. Therefore, since the dimensions orthogonal to
the direction of the boost remain unchanged, one should expect

\begin{equation}\label{}
  R_s^2(K_T) = \langle \tilde{x}_{side}^2 \rangle_0,
\end{equation}
while the dimension perpendicular to the boost gets transformed according to the usual
rules and one should get

\begin{equation}\label{}
  R_o^2(K_T) = \langle \tilde{x}_{out}^2 \rangle_0 -
  2\beta_T\langle \tilde{x}_{out}\tilde{t} \rangle_0 +
  \beta_T^2\langle \tilde{t}^2 \rangle_0.
\end{equation}
The standard recommendation had been to identify the spatial radius in the transverse
direction with $R_s$, to neglect the correlation term $\langle \tilde{x}_{out}
\tilde{t} \rangle_0 $ and to calculate the time span of the production process from
the formula (cf. e.g. \cite{ADL}):

\begin{equation}\label{}
  \tau_0^2 \equiv \langle \tilde{t}^2 \rangle \approx \langle \beta_T^{-2}\rangle
  (R_o^2 - R_s^2).
\end{equation}
This, however, becomes incredible when $R_o \approx R_s$ and absurd when $R_o < R_s$.

An obvious improvement is to take into account the correlation term. The hydrodynamic
models, however, predict

\begin{equation}\label{}
  \langle \tilde{x}_o \tilde{t} \rangle < 0,
\end{equation}
which makes the situation even worse. On the other hand, the result of the
hydrodynamic model has a clear physical interpretation and rejecting it is a serious
decision. The picture behind it is that in the (expanding) cylindrical interaction
volume hadronization begins at the surface and progresses inwards, so that particles
produced at small $\tilde{x}_{out}$ are produced late. The opposite prediction

\begin{equation}\label{}
\langle \tilde{x}_o \tilde{t} \rangle > 0
\end{equation}
occurs in the so-called microscopic models \cite{LKP}, \cite{MOG} where the expansion
according to Euler's hydrodynamics is replaced by an expansion according to some
approximation to Boltzmann's equation. The physical interpretation of this result is
not quite clear yet, but at present the predicted correlation is too weak to explain
the observed decrease of the ratio $R_o(K_T)/R_s(K_T)$. The description of some more
ideas on how to solve this part of the puzzle can be found in the review paper
\cite{KOH}. All these approaches, however, seem to have serious problems.

\section{Approximate equality $\frac{\partial R_s}{\partial K_T} \approx
\frac{\partial R_o}{\partial K_T}$}

Experimentally the transverse momentum dependence of $R_s$ and of $R_o$ is similar --
both drop with increasing $K_T$. Studies of the ratio $R_s(K_T)/R_o(K_T)$ indicate
that $R_s$ drops somewhat faster than $R_o$, Since data does not start at $K_T = 0$,
this does not necessarily mean that $R_o(K_T) < R_s(K_T)$. There could be a rise of
$R_o(K_T)/R_s(K_T)$ in the small $K_T$ region. Nevertheless, these observations
contradict the following simple argument consistent with most models.

Let us consider the case, when $K_T$ is large. Then $\textbf{K}_T$ should be strongly
correlated with the transverse velocity $\textbf{v}_T$ of the corresponding element of
the expanding gas or liquid. It should not be a bad approximation to assume that
roughly the directions of $\textbf{K}_T$ and $\textbf{v}_T$ coincide. Thus, the $out$
components of the particle momenta are generated mostly by the collective velocity
$\textbf{v}_T$ and less by the velocities of the particles in the rest frame of the
element, which should be roughly isotropic, characterized by some temperature $T$. The
side components of the momenta, on the other hand, result mostly from the internal
motion -- thermal in thermodynamic models. This picture implies that $R_o(K_T)$ is
likely to change with $K_T$, while little or no change is expected for $R_s(K_T)$. The
expectation

\begin{equation}\label{}
  \frac{\partial R_s}{\partial K_T} \approx 0,
\end{equation}
which as we now know contradicts experiment, was an important point in Bertsch's
argument in favor of the $L,out,side$ coordinate frame \cite{BER}. It is amazingly
robust and holds not only in hydrodynamic models, where local temperature is a natural
concept, but also in microscopic models \cite{ENO}.

\section{$R_L$ smaller than expected}

It is well known that the parameter $R_L$ measures the size of the "homogeneity
region" and not the total size of the interaction region in the $L$ direction. The
reason is that only pairs of identical particles with similar momenta contribute to
BEC and, because of the strong $\textbf{p} - x$ correlations at production, the
particles with similar momenta cannot be produced very far from each other. Models
typically give

\begin{equation}\label{}
  R_L = \tau_0 \sqrt{\frac{T}{M_T}} f(\frac{T}{M_T}, \frac{K_T}{K_0}),
\end{equation}
where $\tau_0$ is the life time of the interaction region (not to be confused with the
much shorter time interval, where hadronization takes place), $T$ and $M_T$ are
respectively the temperature and the transverse mass of the pair of particles and $f$
is a slowly varying function. The simplest choice -- $f =$ Const -- has been made by
Makhlin and Sinyukov \cite{MAS}. The formula

\begin{equation}\label{}
R_L = \frac{A}{\sqrt{M_T}},
\end{equation}
where $A$ is a constant,  has been successfully used to fit the data \cite{ADC}. The
problem is that the experimental values of the constant $A$, which rises from $2.19
\pm 0.05$ fm GeV$^{1/2}$ at AGS energies ($ \sqrt{s} = 4.1/4.9$ GeV/nucleon) to $3.32
\pm 0.03$ fm GeV$^{1/2}$ at RHIC energies ($\sqrt{s} = (130 - 200)$ GeV/nucleon)
\cite{ADC}, are much smaller than expected \cite{KOH}.

The difficulty can be overcome in a number of ways. One can choose a very small value
of $\tau_0$, which corresponds to expansion with ultrasonic speed \cite{KOH}, or
assume early chemical freeze-out, i.e. no chemical equilibrium among the final
particles, \cite{HIM}. The trouble is that these modifications do not solve the
problem with $R_s$ and spoil the good agreement with experiment for the single
particle distributions  \cite{KOH}. Thus, this problem, though perhaps less striking
than the other two parts of the RHIC puzzle, remains puzzling.

\end{document}